\documentclass{PoS}

\usepackage{subfigure}

\def\comment#1{}

\title{Symanzik flow on HISQ ensembles}

\ShortTitle{Symanzik flow on HISQ ensembles}

\author{A.~Bazavov$^a$, C.~Bernard$^b$, \speaker{N.~Brown}$^b$, C.~DeTar$^c$,  J.~Foley$^c$, Steven~Gottlieb$^d$, U.M.~Heller$^e$, J.E.~Hetrick$^f$, J.~Laiho$^g$
    \hspace{-1.0mm}\thanks{Present address: Department of Physics, Syracuse University, Syracuse, New York,
USA}, L.~Levkova$^{c}$, M.~Oktay$^c$, R.L.~Sugar$^h$, D.~Toussaint$^i$, R.S.~Van~de~Water$^{j}$, R.~Zhou$^d$
\hspace{-1.0mm}\thanks{Present address: Fermi National Accelerator Laboratory, Batavia, IL 60510, USA} \,(MILC Collaboration)\\

$^a$ Physics Department, Brookhaven National Laboratory, Upton, NY 11973, USA\\
$^b$ Department of Physics, Washington University, St. Louis, MO 63130, USA\\
$^c$ Department of Physics and Astronomy, University of Utah, Salt Lake City, UT 84112, USA\\
$^d$ Department of Physics, Indiana University, Bloomington, IN 47405, USA\\
$^e$ American Physical Society, One Research Road, Ridge, NY 11961, USA\\
$^f$ Physics Department, University of the Pacific, Stockton, CA 95211, USA\\
$^g$ SUPA, Department of Physics and Astronomy, University of Glasgow, Glasgow, G12~8QQ, United Kingdom\\
$^h$ Physics Department, University of California, Santa Barbara, CA 93106, USA\\
$^i$ Physics Department, University of Arizona Tucson, AZ 85721, USA\\
$^j$ Theoretical Physics Department, Fermi National Accelerator Laboratory, Batavia 60510, USA\\
E-mail: \email{brownnathan@wustl.edu}
}

\abstract{We report on a scale determination with gradient-flow techniques on the $N_f=2+1+1$ HISQ ensembles generated by the MILC collaboration.  The lattice scale $w_0/a$, originally proposed by the BMW collaboration, is computed using Symanzik flow at four lattice spacings ranging from 0.15 to 0.06 fm. 
With a Taylor series ansatz, the results are simultaneously extrapolated to the continuum and interpolated to physical quark masses.   We give a preliminary determination of the scale $w_0$ in physical units, along with associated systematic errors, and compare with results from other groups.  We also present a first estimate of autocorrelation lengths as a function of flowtime for these ensembles.}

\FullConference{31st International Symposium on Lattice Field Theory LATTICE 2013\\
		 July 29 -- August 3, 2013\\
		 Mainz, Germany}

\begin{document}

\section{Introduction}
\vspace{-2mm}
Scale setting holds central importance in lattice QCD. The precision with which any dimensionful quantity can be computed is limited by the precision with which the scale is determined in physical units. Furthermore, continuum extrapolation of any quantity, dimensionful or dimensionless, requires precise determination of the relative scale between ensembles with different bare couplings. 
Any dimensionful quantity that is finite in the continuum limit may be used for scale-setting, but if
it is not experimentally measured its value in physical units needs to be determined on the lattice
by comparison to
an experimentally determined quantity. 
An ideal scale-setting quantity should be easy and cheap to compute on the lattice, have small statistical errors, and be relatively insensitive to systematic issues, such as finite-volume effects or differences between simulated and physical quark-mass values. 
This has led to the consideration of quantities such as $r_0$ and $r_1$, defined from the static quark potential, and, more recently, $t_0$ \cite{luscher} and $w_0$ \cite{bmw} from gradient flow \cite{luscher_origin}. Gradient flow has received a particularly high interest level over the past year, as evident from the two plenary talks dealing extensively with the flow and its applications \cite{luscherLAT,rainer}. This interest stems from the particular ease with which a scale from the flow can be computed and the resulting small statistical errors. Computing the gradient flow requires no costly quark propagator computations and avoids fits, such as to the asymptotic time behavior of correlation functions or Wilson loops.

Here, we present our computation of $w_0/a$ on the MILC HISQ ensembles, and perform a continuum extrapolation to determine $w_0$ in physical units.  We also investigate the autocorrelation length of the
energy density as a function of flow time on these ensembles.


\vspace{-1mm}
\section{Gradient Flow and the Scale $w_0$}
\vspace{-2mm}
	Gradient flow \cite{luscher_origin}  is a smoothing of the original gauge fields $U$ towards stationary points of the gauge action $S$. Successive links $V(t)$ are updated according to the diffusion equation,
	\[ \frac{d}{dt} V(t)_{i,\mu} = - V_{i,\mu} \frac{\partial S(V)}{\partial V_{i,\mu}}\,, \hspace{3mm} V(t)_{i,\mu}(0) = U_{i,\mu} 
	\hspace{15mm} \left[ \frac{dA_{\mu}}{dt} = D_{\nu} F_{\nu\mu} \right]\;, \]
where the equation in square brackets is the corresponding continuum flow equation. The flow time $t$ increases as the gauge fields are smoothed. By repeatedly integrating infinitesimal smearing steps, the diffusion equation can be solved numerically. Note that the action $S$ is not necessarily the same as the action used to generate the original gauge fields. As the gauge fields diffuse, high momentum artifacts of the lattice spacing are removed. Thus, statistical fluctuations and discretization effects are suppressed while preserving low-energy physics.
 
To extract the scale using the gradient flow one needs a dimensionful quantity. One of the simplest is the flow time $t$ which has units of $a^2$. Pick a reference timescale $t_0$ at which a dimensionless quantity reaches a predefined value. If this quantity is finite in the continuum limit, then $t_0/a^2$ will be independent of lattice spacing up to discretization corrections in powers of $a$. L\"uscher  \cite{luscher}  has shown the energy density $\langle E(t) \rangle$ is finite to next-to-leading order (when expressed in terms of renormalized quantities), where $\langle E(t) \rangle = \langle G_{\mu\nu}^a(t) G_{\mu\nu}^a(t) \rangle / 4$ and $G_{\mu\nu}^a(t)$ is the clover-leaf definition of the field strength tensor at flow time $t$. The reference timescale $t_0$ is chosen to satisfy $t_0^2 \langle E(t_0) \rangle = 0.3$. Based on empirical evidence, the discretization effects can be reduced by considering the slope
	$W(t) = t \frac{d}{dt} \left( t^2 \langle E(t) \rangle  \right)$
and defining the new reference timescale $w_0$ where $W(w_0^2) = 0.3$ \cite{bmw}. In both cases, the cutoff of $0.3$ is chosen to minimize discretization effects from small flow times and finite volume effects at large flow times.

\vspace{-1mm}	
\section{HISQ Ensemble Results}
\vspace{-2mm}
We have computed the scale $w_0/a$ on the MILC $N_f=2+1+1$ HISQ ensembles \cite{milcHISQ}. The tree-level Symanzik improved action is used in the flow, partially for comparison to earlier work with Wilson flow on the same ensembles by HPQCD \cite{hpqcd}. Note that the energy density is computed with the clover-leaf definition of the field strength tensor; thus, the final scale $w_0$ is not Symanzik improved. Numerical integration is performed with the fourth-order Runga-Kutta scheme originally proposed in 
Ref.~\cite{luscher}. We found a step size of $\epsilon=0.03$ sufficient to render integration step-size errors negligible. Results for each ensemble are compiled in Table 1.

\begin{table}
\label{hisqtables}
\begin{center}
	\subfigure[Physical strange mass ensembles ($m'_s=m_s$)]{%
		\label{phystable}
		\begin{tabular}{|l|l|l||l|l|}
			\hline
			$\approx a$(fm) & $m_l/m_s$ & volume & $N_{run}$ & $w_0/a$ (stat) [\%]\\
			\hline
			\hline
			{\small 0.15}	 & 0.2 & $16^3 \times 48$ & 1021 & 1.1221 (06) [0.06\%] \\
			0.15	 & 0.1 & $24^3 \times 48$ & 1000 & 1.1381  (04) [0.04\%] \\
			0.15	 & 0.037 & $32^3 \times 48$ & 999 & 1.1468 (03) [0.03\%] \\
			\hline
			0.12 & 0.2 & $24^3 \times 64$ & 1040 & 1.3835  (07) [0.05\%] \\
			0.12 & 0.1 & $24^3 \times 64$ & 1020 & 1.4020  (10) [0.07\%] \\
			0.12	 & 0.1 & $32^3 \times 64$ & 999 & 1.4047 (06) [0.05\%] \\
			0.12 & 0.1 & $40^3 \times 64$ & 1001 & 1.4041 (04) [0.03\%] \\
			0.12	 & 0.037 & $48^3 \times 64$ & 34 & 1.4168  (10) [0.07\%] \\
			\hline
			0.09	 & 0.2 & $32^3 \times 96$ & 102 & 1.8957 (16) [0.08\%] \\
			0.09	 & 0.1 & $48^3 \times 96$ & 151 & 1.9296  (09) [0.05\%] \\
			0.09	 & 0.037 & $64^3 \times 96$ & 53 & 1.9473 (11) [0.06\%] \\
			\hline
			0.06	 & 0.2 & $48^3 \times 144$ & 127 & 2.8956 (26) [0.09\%] \\
			0.06	 & 0.1 & $64^3 \times 144$ & 46 & 2.9486  (31) [0.11\%] \\
			0.06	 & 0.037 & $96^3 \times 192$ & 49 & 3.0119 (18) [0.06\%] \\
			\hline
		\end{tabular}
	}
	\subfigure[Non-physical strange mass ensembles ($m_s'\not=m_s$)]{%
		\label{mstable}
		\begin{tabular}{|l|l|l|l||l|l|}
			\hline
			$\approx a$(fm) & $m_l/m_s$ & $m_s'/m_s$ & $volume$ & $N_{run}$ & $w_0/a$ (stat) [\%]\\
			\hline
			\hline
			0.12 & 0.10  & 0.10 & $32^3 \times 64$ & 102 & 1.4833 (13) [0.09\%]\\
			0.12 & 0.10  & 0.25 & $32^3 \times 64$ & 204 & 1.4676 (11) [0.07\%]\\
			0.12 & 0.10	 & 0.45 & $32^3 \times 64$ & 205 & 1.4470 (11) [0.08\%]\\
			0.12 & 0.10	 & 0.60 & $32^3 \times 64$ & 107 & 1.4351 (20) [0.14\%]\\
			\hline
			0.12 & 0.175 & 0.45 & $32^3 \times 64$ & 134 & 1.4349 (13) [0.09\%]\\
			0.12 & 0.20  & 0.60 & $24^3 \times 64$ & 255 & 1.4170 (10) [0.07\%]\\
			0.12 & 0.25  & 0.25 & $24^3 \times 64$ & 255 & 1.4336 (16) [0.11\%]\\
			\hline
		\end{tabular}
	}
\end{center}
\vspace{-5mm}
\caption{Results for $w_0/a$. $N_{run}$ is the number of configurations
analyzed; $m_s$ is the (approximate) physical strange quark mass, while $m'_s$ is the mass value for the ensemble.
The  $m'_s=m_s$, $a\!\approx\!0.12$ fm, $m_l/m_s=0.1$, $24^3 \times 64$ ensemble is not included in continuum extrapolations due to large finite volume effects.
\label{tables}
\vspace{-3mm}
}
\end{table}

\begin{figure}
\begin{center}
	\subfigure{%
		\label{relerror}
		\includegraphics[width=0.47\textwidth]{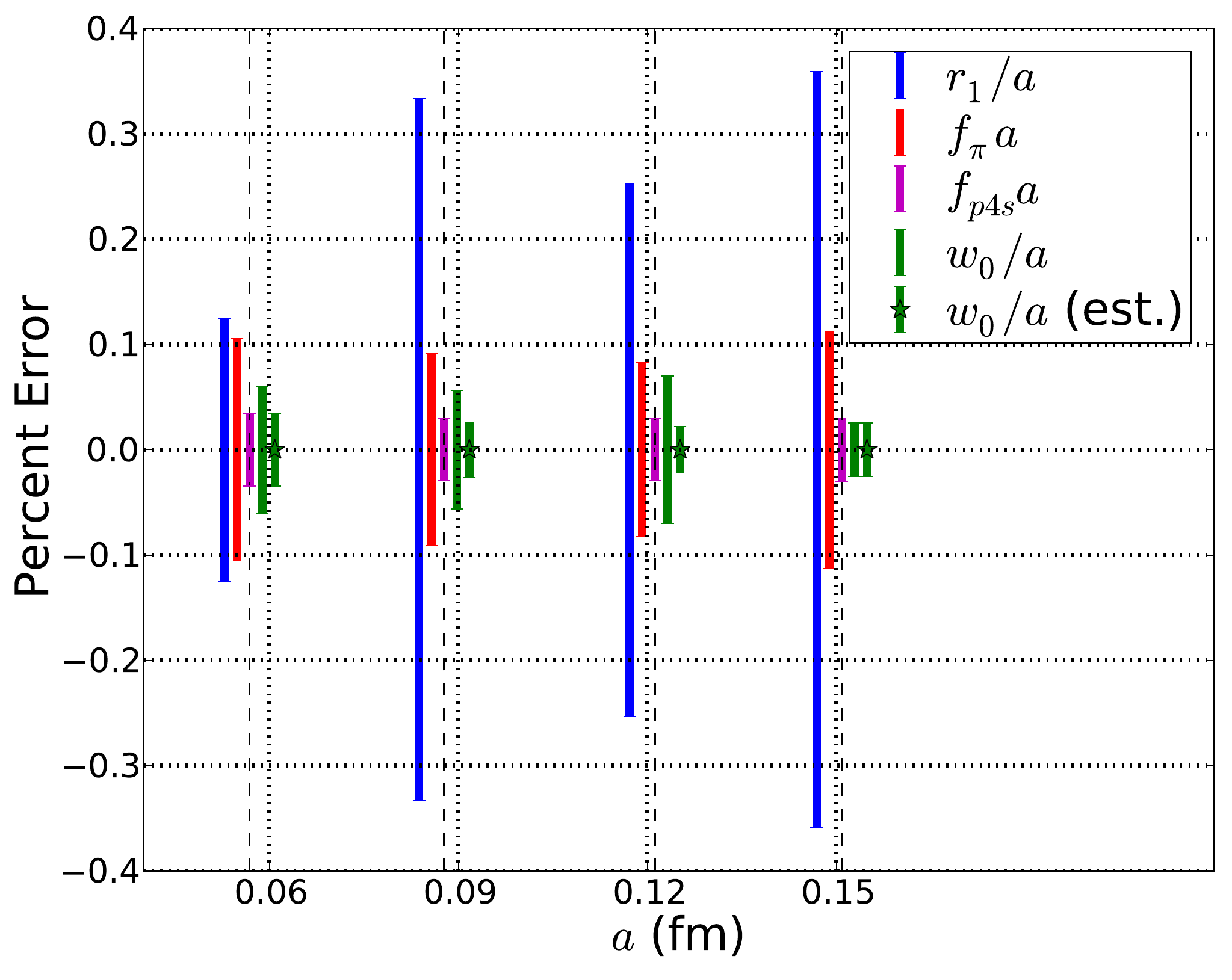}
	}%
	\subfigure{%
		\label{naivefit}
		\includegraphics[width=0.49\textwidth]{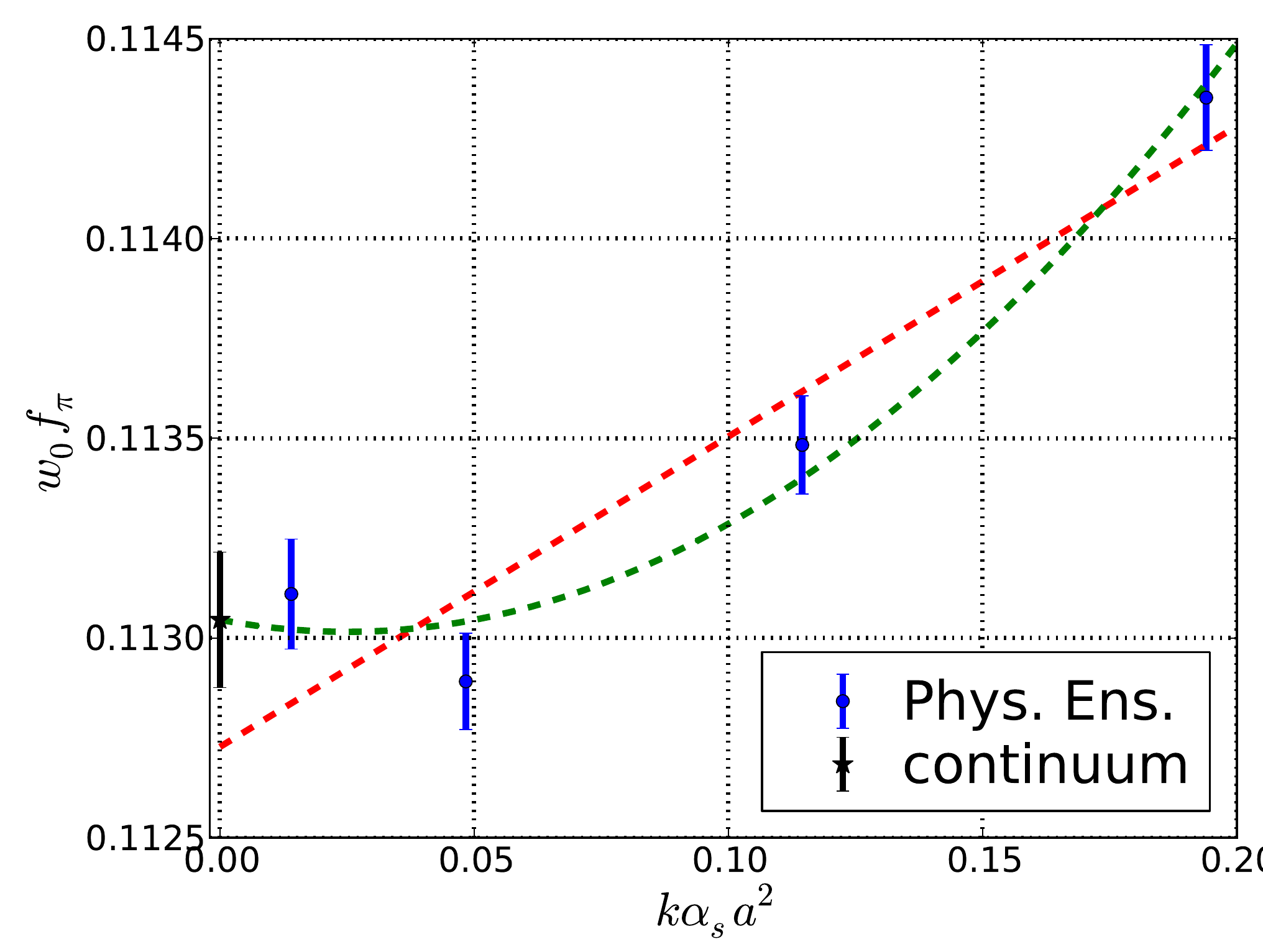}
	}
\end{center}
\vspace{-5mm}
\caption{ Left) Comparison of percent error for different scale setting quantities. The dashed lines denote the lattice spacing and all values are for the physical quark-mass ensembles. The star  points for $w_0$ are estimates of the percent error 
obtainable for the full ensembles.
Right) Simple linear and quadratic extrapolations are plotted. No adjustments are made for mistunings of quark masses. The central value and statistical error are represented by the black, continuum point.
\label{simpleresults}
\vspace{-3mm}
}
\end{figure}

As can be seen in Table 1\comment{\subref*{phystable} and \subref*{mstable}}, the statistical error of $w_0/a$ is below 0.1\% for almost all ensembles. Furthermore, the gradient flow has been run on only a fraction of the configurations in most ensembles. Each ensemble has or will have approximately 1000 configurations, which leaves plenty of room for reduction of statistical errors. Note, the reduction would be particularly large for the finer $a=0.09$ and $0.06$ fm ensembles where $N_{run}$ is still small. As a comparison, Fig.~1 (Left)\comment{\subref*{relerror}} plots the percent error of $w_0/a$ and other scale-setting quantities against $a$ for the four physical quark mass ensembles.
Only $f_{p4s}a$, the pseudoscalar decay constant at the fiducial point with valence-quark masses 0.4 times the strange quark mass \cite{milcHISQ}, has lower percent errors than $w_0/a$. However, $f_{p4s}a$ was computed over the entirety of each ensemble. Using conservative estimates of the autocorrelation length, the statistical error of $w_0/a$ should be comparable to or better than the error of $f_{p4s}a$ if computed on the same number of configurations.

A simple continuum extrapolation can be performed by including only the physical quark-mass ensembles. With just these ensembles, quark mass mistuning effects cannot be accounted for, and the statistical error will be larger than from a fit to the complete data set. Nevertheless, this extrapolation provides a double check on the final value from the more complicated fit. The results of a linear and quadratic fit to $w_0 f_\pi$ in powers of $\alpha_s a^2$ are shown in Fig.~1 (Right)\comment{\subref*{naivefit}}. Taking $w_0$ from the quadratic fit as the central value, and the difference between the two fits as extrapolation error, this fit gives $w_0 = 0.1711(3)(3)$ fm.

\vspace{-1mm}
\section{Fitting and Extrapolation}
\vspace{-2mm}
	The value of $w_0 f_\pi$ may be simultaneously interpolated to physical quark masses and extrapolated to the continuum by fits to all of our data.  All fit functions used are power series in $a^2$ or $\alpha_s a^2$, $M_\pi^2$, and $2M_K^2-M_\pi^2$ where the latter two are proxies for $m_l$ and $m_s$, respectively.  Starting at linear order, fit functions up to cubic order in each variable (including cross terms) are considered. The lowest order discretization term is always $\alpha_s a^2$, and higher orders are powers of either $a^2$ or $\alpha_s a^2$. Including the non-physical $m'_s$ ensembles gives us the most data to work with and helps correct mistuning errors. However, those ensembles cover a wide range of $m'_s$ values, so 
we also consider fits that drop one or more of those ensembles with the lightest strange-quark masses. 
Five different cutoffs on $m'_s$ were considered. In total, there are five choices for parameterizing discretization effects (powers of $a^2$ or $\alpha_s a^2$), three possible orders of $M_\pi^2$ and $2M_K^2-M_\pi^2$, and five different datasets to choose from (corresponding to different cutoffs on $m'_s$), yielding 215 different fits. Note that 10 such fits would have more parameters than degrees of freedom and are therefore excluded from the count.
	
\begin{figure}
\begin{center}
	\subfigure{%
		\label{combfit}
		\includegraphics[width=0.53\textwidth]{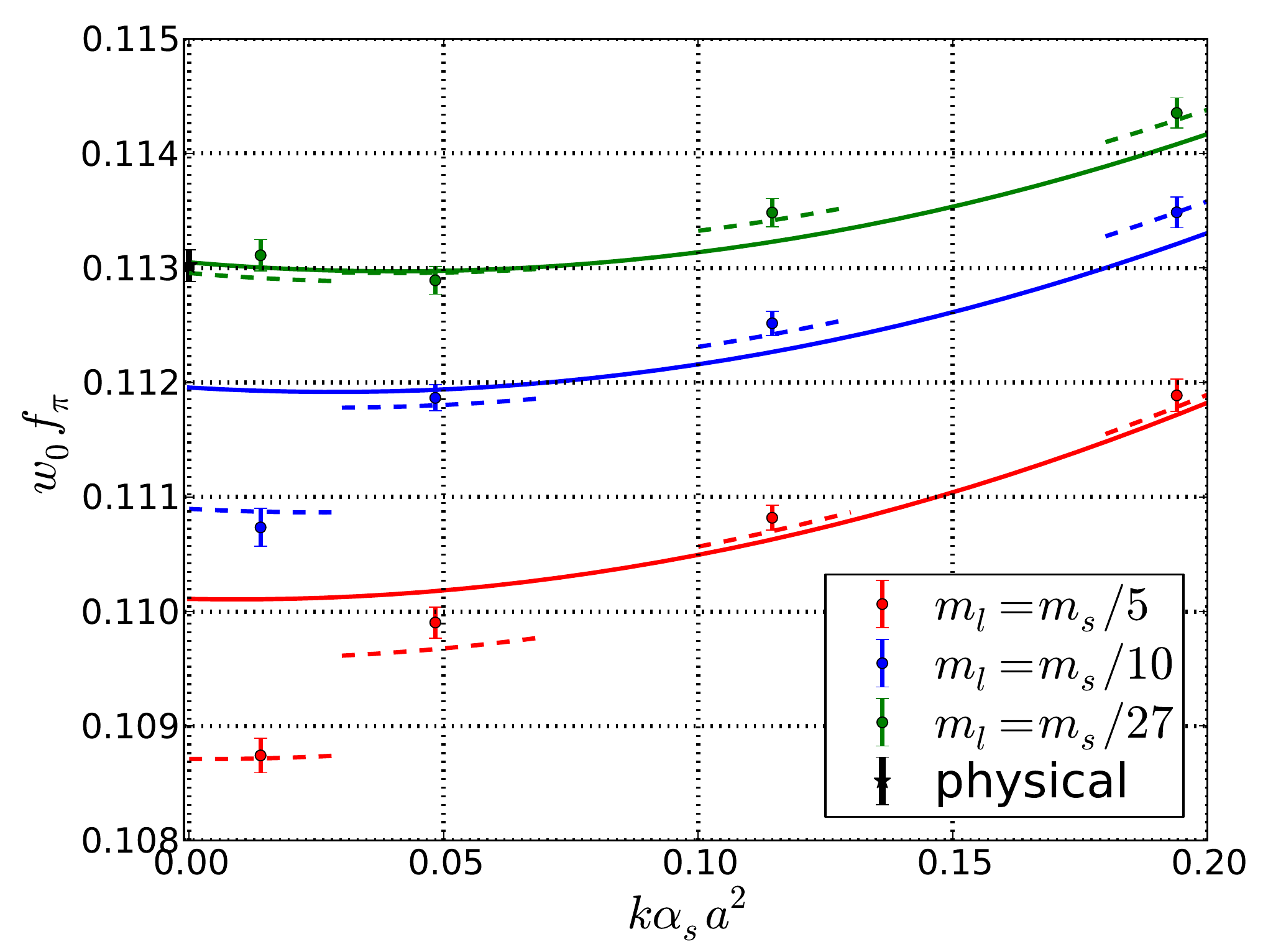}
		}%
	\subfigure{%
		\label{histfit}
		\includegraphics[width=0.45\textwidth]{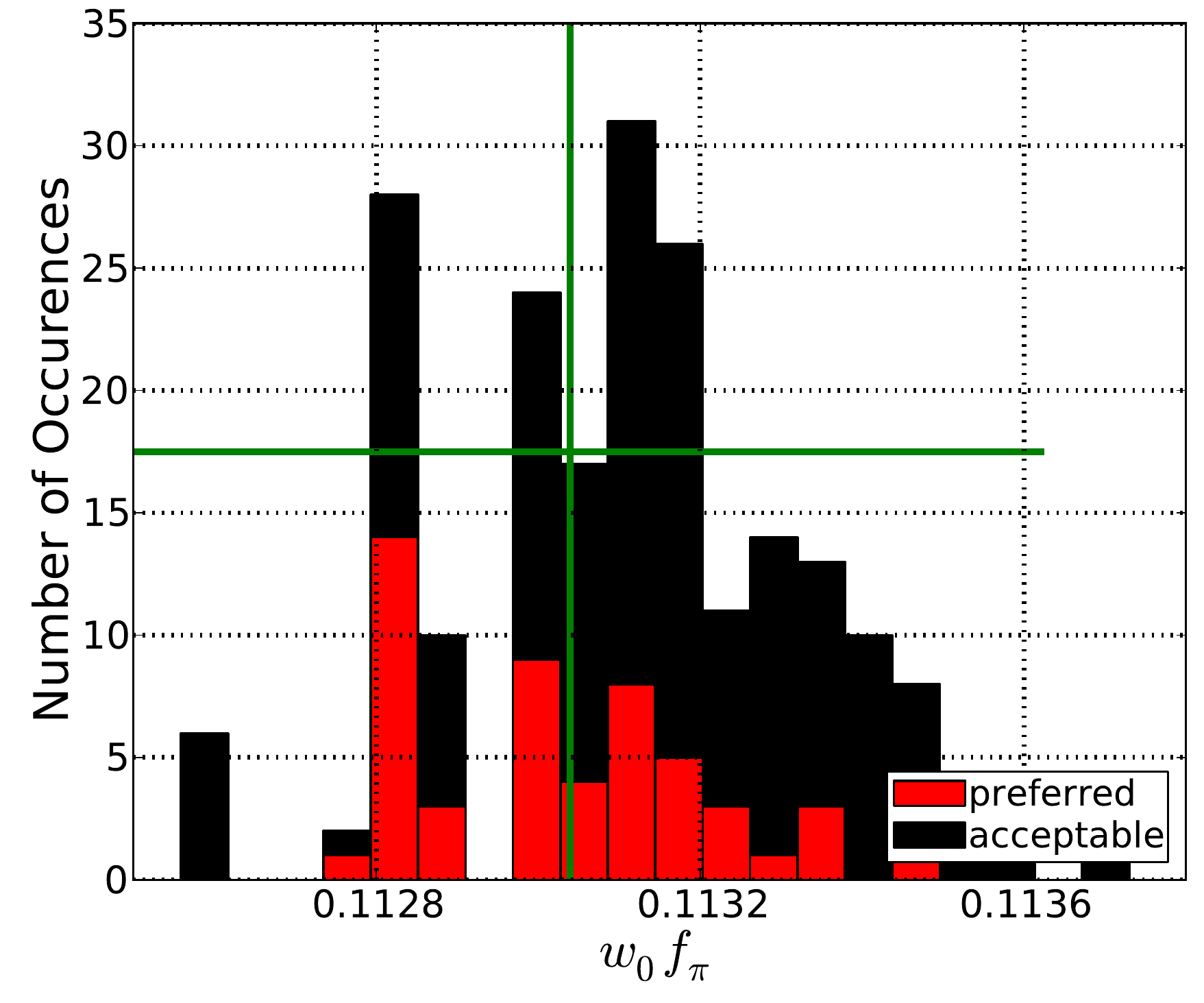}
	}
\end{center}
\vspace{-5mm}
\caption{
Left) A representative best fit used to compute $w_0$ at physical quark mass and extrapolated to the continuum. Only $m'_s=m_s$ ensembles are plotted, but the fit includes all $m'_s \le m_s$ ensembles. Dotted lines represent the fit through each ensembles quark masses, while the solid lines are for retuned quark masses per legend.
 Right) A histogram of the continuum, physical quark mass $w_0$ for the acceptable fits (p-value greater than 0.01). 
Preferred fits (see text) are  plotted in red. The green cross represents the chosen central value and extrapolation error.
\label{fitresults}
\vspace{-3mm}
}
\end{figure}

A fit is considered acceptable if it has a p-value greater than 0.01. An acceptable fit is considered to 
be ``preferred'' if it displays at least two of the following three attributes: a p-value greater than 0.1, at least 
twice as many degrees of freedom as fit parameters, or a deviation smaller than one sigma from the 
physical 0.06 fm ensemble, which is the most important ensemble since it has the finest lattice spacing and
physical masses.
A representative preferred fit is plotted in Fig.~2 (Left)\comment{\subref*{combfit}}. The values of $w_0 f_\pi$ for 
all acceptable fits are binned and placed in the histogram in Fig.~2 (Right)\comment{\subref*{histfit}}.  From
among the preferred fits, a central fit is chosen close to the median of the preferred and acceptable fits. The error from this procedure is conservatively 
estimated as the width of the histogram. The chosen central fit has the functional form
\[ w_0f_\pi = c_0 + c_1 \alpha_s a^2 + c_2 M_\pi^2 + c_3 (2 M_K^2-M_\pi^2) + c_4 a^4 + c_5 M_\pi^4 + c_6 a^2 M_\pi^2 \]
with 20 points fit (all non-physical $m_s$ ensembles included). The fit has $\chi^2/dof = 18.6/13$, $p=0.14$, and 
is $1.4\sigma$ from the re-tuned physical 0.06 fm ensemble.

\vspace{-1mm}
\section{Autocorrelations}
\begin{figure}
\begin{center}
	\subfigure{%
		\label{avglengths}
		\includegraphics[width=0.51\textwidth]{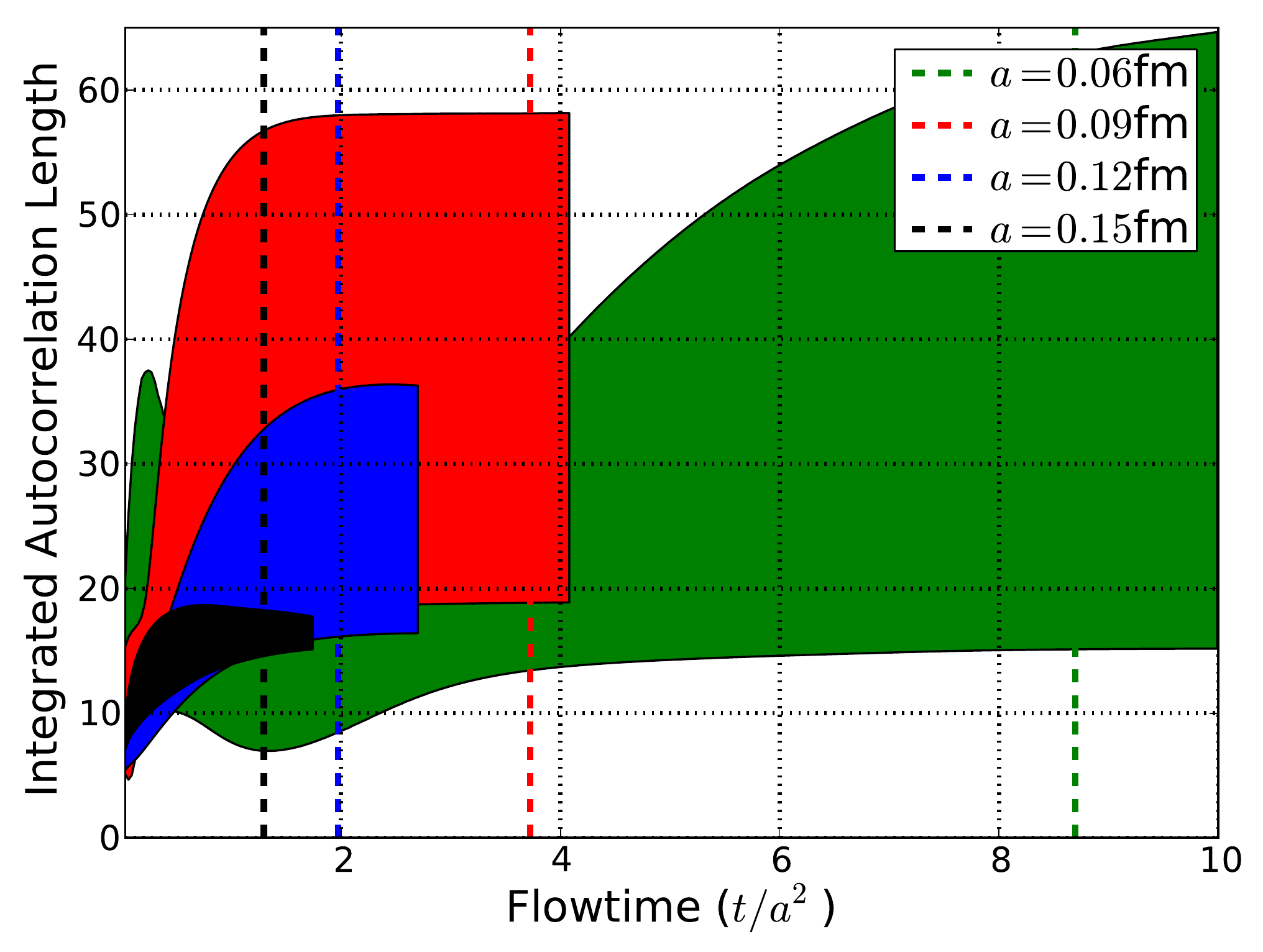}
	}%
	\subfigure{%
		\label{interror}
		\includegraphics[width=0.48\textwidth]{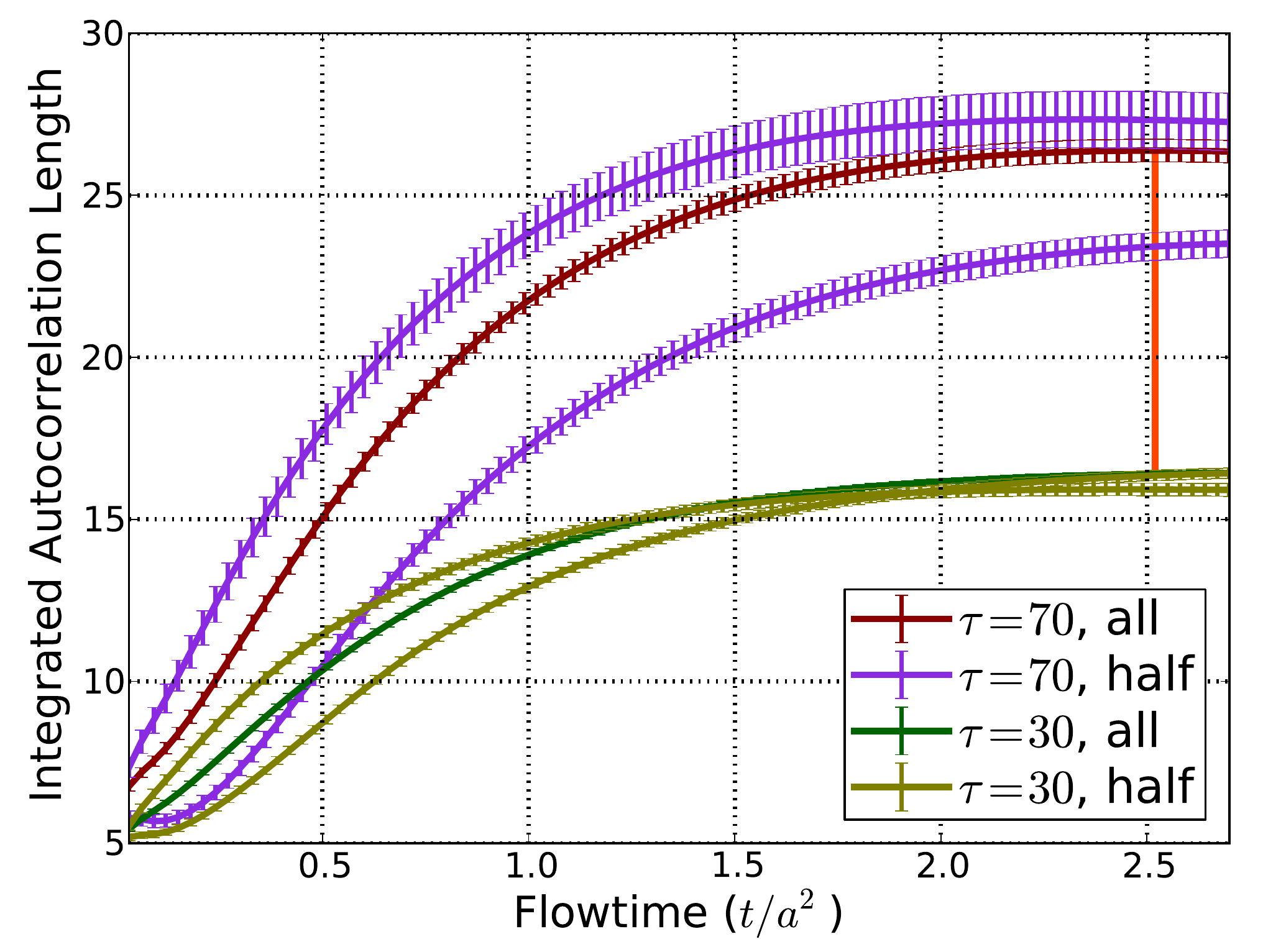}
	}
\end{center}
\vspace{-5mm}
\caption{Left) The integrated autocorrelation length as a function of flowtime for ensembles with $m_l/m_s=0.1$. The plotted error bars are a combination of the statistical error and variation with the choice of integration region, added in quadrature. Dashed vertical lines denote the flow time that determines $w_0$.
Right) The jackknifed, integrated autocorrelation length for different choices of the cutoff $\tau$, on the 0.12fm $m_l/m_s=0.1$ ensemble. The plots correspond to the largest and smallest integration length where the statistical errors are not completely uncontrolled. The autocorrelation lengths for the two halves of the ensemble are plotted for each cutoff. The error associated with the cutoff is denoted by the orange vertical line.
\label{autocorrelations}
\vspace{-4mm}
}
\end{figure}

\vspace{-2mm}
To help pick the jackknife bin size and control correlations, we compute the autocorrelation function of $
\langle E(t) \rangle$ as a function of the flow time $t$ and the number, $\tau$, of unit trajectories separating 
configurations. Autocorrelation functions are notoriously difficult to estimate precisely, and for many of the 
ensembles the $\tau$ between the configurations used to determine $w_0$ is not small enough to see 
correlations reliably. In particular, at 0.09 and 0.06 fm, running over the entire ensembles to compute 
autocorrelation lengths is  computationally expensive.  Instead, we run on only 50 and 25 equilibrated 
configurations separated by 24 unit trajectories from the $m_l/m_s=0.1$, $a=0.09$ and 0.06 fm ensembles, 
respectively. For the ensembles at 0.15 and 0.12 fm where the full ensembles have been run, we have a 
much better estimate of the autocorrelation function.  A comparison of the integrated autocorrelation length 
as a function of flow time for different lattice spacings and light sea quark masses is plotted in 
Fig.~3 (Left)\comment{\subref*{avglengths}}.

To estimate the statistical error on the integrated autocorrelation length, we jackknife the results for $\langle E(t) \rangle$ and compute the average autocorrelation function across all flow times. We integrate the
the average autocorrelation function up to a cutoff value of $\tau$ where the autocorrelations become indistinguishable from noise. 
There is additional error from the choice of the cutoff, which is not completely separable from the estimate of the statistical error. We attempt to account for this error by measuring the difference between the largest and smallest integrated autocorrelation lengths for which the statistical errors are not completely uncontrolled.  As
a function of increasing cutoff, the errors are considered  to have become
``uncontrolled'' when the deviation between the result on the full dataset and those on the first or second halves is more than 3 sigma.  Fig.~3 (Right)\comment{\subref* {interror}} illustrates the integration error estimation process for the 0.12 fm $m_l/m_s=0.1$ ensemble. The integration region yielding the largest integrated autocorrelation length is chosen for each ensemble in Fig.~3 (Left)\comment{\subref*{avglengths}} in order to estimate the worse-case scenario.

\vspace{-1mm}
\section{Results and Conclusion}
\vspace{-2mm}
For each HISQ ensemble, we jackknife $w_0/a$ with a bin size larger than the estimated autocorrelation length. Then we compute $w_0 f_\pi$ through a simultaneous extrapolation to the continuum and physical quark masses, and use the experimental value of $f_{\pi}$ to yield $w_0$ in physical units. This yields the final (but still preliminary) value of 
\[w_0 = 0.1711 (2)(8)(2)(3)\; {\rm fm}\;,
\] 
where the errors are statistical, systematic error from the continuum extrapolation and chiral interpolation, residual finite volume error in $a f_\pi$, and experimental error in $f_\pi$, respectively. The result is in agreement with simple continuum extrapolation through the physical mass ensembles, which gives $w_0 = 0.1711 (3)(3)$ fm (statistical and rough continuum extrapolation errors only).

HPQCD has also calculated $w_0$ on a the 0.15 to 0.09 fm, physical strange quark mass MILC HISQ 
ensembles \cite{hpqcd}, a subset of the ensembles considered here. They used the Wilson action in the 
flow, and also set the absolute scale with $f_\pi$, as we have done here. They report a final value of $w_0 = 
0.1715(9)$ fm, well within one standard deviation of our result.  At the moment, there are no other complete 
computations of $w_0$ with $N_f=2+1+1$ fermions. ETM reported a preliminary value of $w_0 = 0.1782$ 
fm on their $N_f=2+1+1$ twisted mass ensembles, but the 
error analysis at that time was not complete \cite{etm}. BMW reports $w_0 = 0.1755(18)(04)$ fm on $N_f=2+1$ 2-HEX 
smeared Wilson-clover ensembles,  a value about 2.1 (joint) sigma from ours. It is conceivable that the difference between their result and ours is due to the different numbers
of flavors 
in the sea. However, as reported in Ref.~\cite{rainer}, there is no obvious trend in $w_0$ as $N_f$ increases.

\vspace{-2mm}
\acknowledgments
\vspace{-2mm}
This work was supported by the U.S. Department of Energy and the National Science Foundation. Computations for the calculation of Symanzik flow were carried out with resources provided by the Texas Advanced Computing Center (TACC).  The HISQ gauge configurations were generated with resources provided by the Argonne Leadership Computing Facility, the Blue Waters Project at the National Center for Supercomputing Applications (NCSA), the National Energy Resources Supercomputing Center (NERSC), the National Institute for Computational Sciences (NICS), TACC, the National Center for Atmospheric Research (UCAR), and the USQCD facilities at Fermilab, under grants from the DOE and NSF.

\end{document}